\documentclass[conference]{IEEEtran}
\IEEEoverridecommandlockouts
\usepackage{times}
\usepackage{epsfig}
\usepackage{graphicx}
\usepackage{amsmath}
\usepackage{amssymb}
\usepackage{multicol,multirow}
\usepackage[tableposition=top]{caption}
\usepackage{cite}
\usepackage{booktabs}
\usepackage{bigstrut}
\usepackage{floatrow}
\usepackage{rotating}
\usepackage{adjustbox}
\usepackage{amsmath,url}

\def\BibTeX{{\rm B\kern-.05em{\sc i\kern-.025em b}\kern-.08em
    T\kern-.1667em\lower.7ex\hbox{E}\kern-.125emX}}
\begin{document}

\title{RC-Net: A Convolutional Neural Network for Retinal Vessel Segmentation\\
}
\author{Tariq M. Khan$^1$ \hspace{2cm}
Antonio Robles-Kelly$^1$ \hspace{2cm}
Syed S. Naqvi$^2$\\
$^1$ School of IT, Faculty of Sci. Eng. \& Built Env., Deakin University, Waurn Ponds, VIC 3216, Australia\\
$^2$ Dept. of Electrical and Computer Eng., COMSATS University Islamabad,
Islamabad, Pakistan}



\maketitle

\begin{abstract}
Over recent years, increasingly complex approaches based on sophisticated convolutional neural network architectures have been slowly pushing performance on well-established benchmark datasets. In this paper, we take a step back to examine the real need for such complexity. We present RC-Net, a fully convolutional network, where the number of filters per layer is optimized to reduce feature overlapping and complexity. We also used skip connections to keep spatial information loss to a minimum by keeping the number of pooling operations in the network to a minimum. Two publicly available retinal vessel segmentation datasets were used in our experiments. In our experiments, RC-Net is quite competitive, outperforming alternatives vessels segmentation methods with two or even three orders of magnitude less trainable parameters.
\end{abstract}

\begin{IEEEkeywords}
Medical Image Segmentation, Convolutional Neural Networks, Residual Connections.
\end{IEEEkeywords}

\section{Introduction}

\IEEEPARstart{D}{iabetic} Retinopathy (DR) has recently received a lot of attention due to the pathology's link to long-term diabetes, which is one of the leading causes of preventable blindness worldwide\cite{Khawaja2019a}. Furthermore, DR is one of the leading causes of vision loss, particularly among the working age population. Lesions, a broad term that encompasses microaneurysms (MA), exudates (hard and soft), inter-retinal microvascular abnormalities, hemorrhages (dot and blot), and leakages, are the first signs/symptoms of DR \cite{Soomro2018, 9360745}. As a result, the number and type of lesions that develop on the retina's surface determine the disease's condition and diagnosis. As a result, the accuracy of segmentation of retinal blood vessels, optic cup/disc, and retinal lesions is expected to be critical to the performance of an automated system for large-scale public screening \cite{diagnostics11010114,khan2019use}. Along these lines, the detection of retinal blood vessels has long been regarded as the most difficult challenge, and it is frequently regarded as the most important component of an automated diagnostic system \cite{Khawaja2019a, imtiaz2021screening}. This is due to the tortuosity, density, diameter, branching, and shape of the vessels in the retina, which make them difficult to detect. The centerline reflex, as well as the numerous constituent parts of the retina, such as the optic cup/disc, macula, hard exudates, soft exudates, and so on, which may exhibit lesions or imperfections, make detection even more difficult. Finally, variability in the imaging process can be introduced by the acquisition process and camera calibration parameters.

Methods for detecting blood vessels are supervised in nature, with a machine learning model being trained on a database of manually segmented images \cite{10.1007/978-3-030-63820-7_18}. These methods have been applied to the detection of retinal vessels for diagnosing critical diseases such as glaucoma \cite{Thakoor2019}, diabetic retinopathy (DR) \cite{Zeng2019}, retinal vascular occlusions \cite{Muraoka2013}, age-related macular degeneration (AMD) \cite{Cicinelli2018} and chronic systematic hypoxemia \cite{Traustason2011}. Furthermore, deep learning-based methods have achieved state-of-the-art accuracy in tasks like optic cup/disc detection and vessel detection \cite{Jiang2019a}. As a result, supervised machine learning models have become the preferred method for developing retinal diagnostics systems \cite{khan2020exploiting, khan2020shallow}. Despite their great success, detecting blood vessels in the presence of significant contrast variations and lesions remains a difficult task. When the diameter of the vessels is small, it becomes even more difficult.


Despite the fact that these methods produce supervised segmentation results that are superior to their unsupervised counterparts, training and testing these networks can be time consuming. This is made even more difficult by the scarcity of densely annotated data for a wide range of conditions and imaging modalities. As highlighted by \cite{chen19closerfewshot,musgrave2020metric}, for most segmentation techniques, the use of complex CNN architectures does not yield the best results.


Remember that the number of trainable parameters is heavily influenced by the number of hidden layers as well as the number of filters used in each layer.   
Shallow networks are frequently proposed as an alternative to deep networks in these situations \cite{9207668}. The number of filters used per layer in these shallow networks is also lower than in their deep counterparts.

The purpose of our network design is to make use of the maximum number of filters in each layer in order to reduce their overall complexity. With a large number of filters in a convolution layer, the performance does not improve (if an image with feature variation less) but the complexity increases. The complexity of convolution networks has been reduced in the literature by suggesting low networks with a small number of layers \cite{Howard_2019_ICCV, Ma_2018_ECCV}. Although the number of filters per layer has been reduced by search, this has stemmed from hardware needs \cite{Howard_2019_ICCV}. Moreover, the importance of this in terms of complexity and performance was not discussed in \cite{Howard_2019_ICCV}. Here the number of filters is selected based on the characteristic complexity (a more generic parameter than hardware requirement).


\section{Contribution}

This paper presents a simple yet effective tiny neural network architecture which we call RC-Net.   
This is due to the small number of parameters in RC-Net, which requires a much smaller memory and GPU footprint for testing as compared to alternatives with much larger number of parameters.
Further, RC-Net is  designed to guarantee dual-stream information flow both inside as well as outside of the encoder-decoder pair. To compensate spatial loss created by the fully convolutional layers, internal residual skip connections have been used. In this manner, the outer skip residual connections used here provide direct spatial information from the encoder layer to the decoder. Moreover, only two max-pooling operations are used in the encoder. We have kept the application of pooling to a minimum so as to mitigate the loss in spatial resolution.

To illustrate the utility of our method for purposes of medical image segmentation, we have performed experiments in three different medical applications. These are retinal vessel segmentation, intestinal polyp detection and skin lesion segmentation. Furthermore, in our experiments, our method outperformed many of the alternatives and was always comparable to all the alternatives under consideration, some of which have more than two or three orders of magnitude more trainable parameters.

\section{RC-Net}

\begin{figure*}[!h]
  \centering
  \vspace{-4.5cm}
  \includegraphics[scale=0.19]{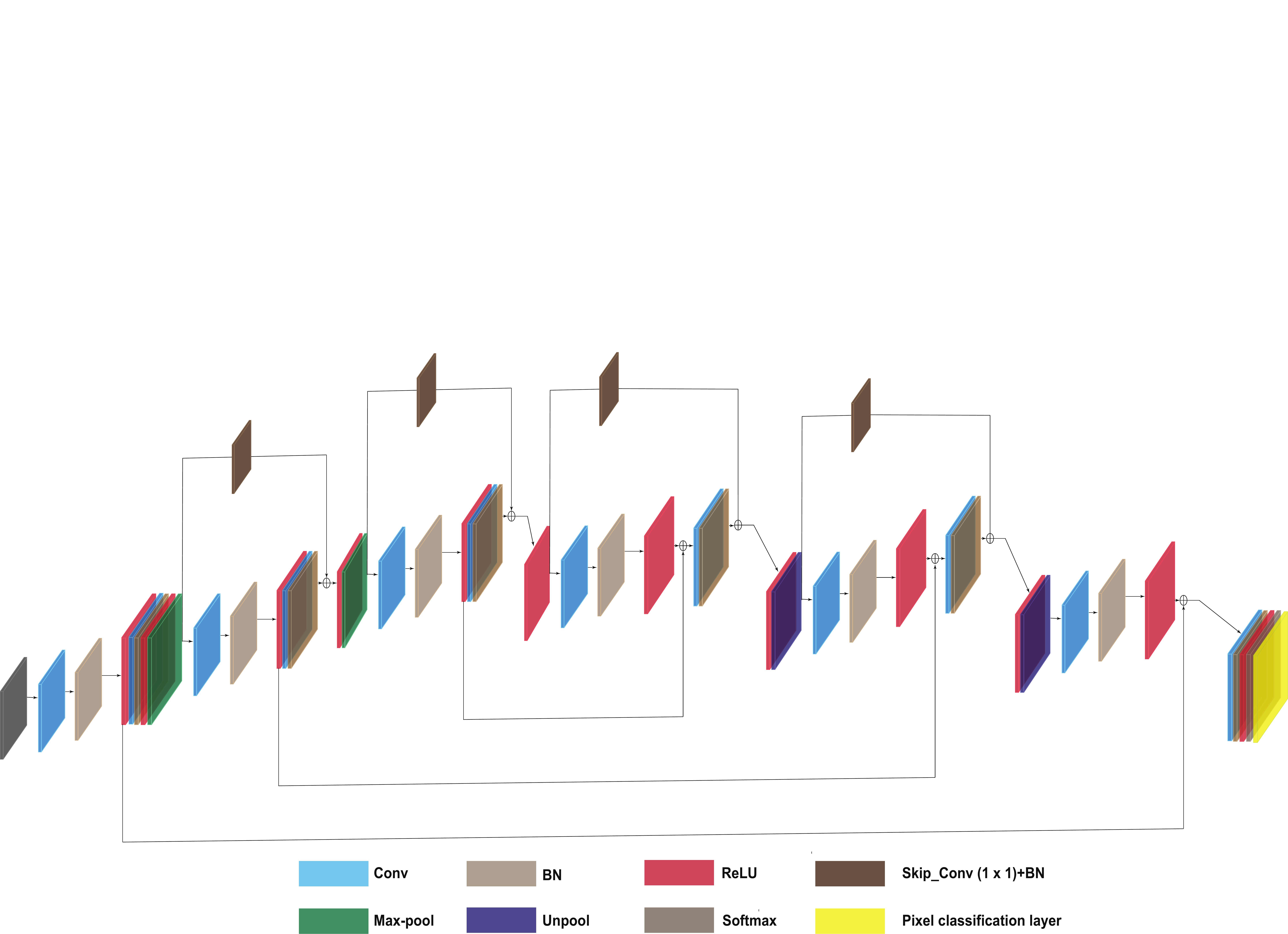}\\
 \caption{Block diagram of the proposed method}\label{newmethodblockdiagram}
\end{figure*}

In this section we present our network and motivate its architecture. In Figure \ref{newmethodblockdiagram} we show the overall architecture of RC-Net. Note that our network has six convolutional blocks, where the first block of these is the input one, followed by two down-sampling convolutional blocks. There is an intermediate convolutional block which bridges between the down and up-sampling blocks. There are two up-sampling convolutional blocks followed by the final convolutional output block which is equipped with the essential final layers responsible of creating the pixel-wise segmentation map.
All the encoder blocks generate the respective collection of features making use of convolutions between the input feature maps and the filter banks. Following \cite{Ioffe2015} we have performed batch normalisation on these features followed by the application of a ReLU. The resulting feature maps are then passed on to the max-pooling or unpooling layers, depending upon whether the block under consideration is a down-sampling or up-sampling one. All max-pooling and unpooling layers have set to be a size of 2$\times$2, non-overlapping with a stride of size 2. In our network, we have used only two max-pooling and two unpooling layers. All the convolutional filters in our network are of size $3 \times 3$.

Note that
in our architecture, the feature maps yielded by the down-sampling blocks are sparse in nature after the unpooling operation. The convolutional filter banks have the effect of ``densifying'' these maps in the up-sampling blocks. These dense feature maps are then normalised by using batch normalisation. The size of the feature maps yielded by the up-sampling blocks are identical to those obtained by the respective down-sampling blocks. The final classification layer is a binary one right after a soft-max classification layer.

Note the network architecture responds to a number of motivations. Firstly, in our architecture we have aimed at using the least possible number of pooling layers. This is since these often reduce the dimension of the feature maps and can also  cause the loss of spatial information. Secondly, we have employed a small number of convolutional layers. Thirdly, we have also reduced the overall  number of convolutional filters within each layer.  Finally, we have profited from fine-grained information making use of residual skip paths.

Thus, in our network, we preserve the boundary structure of the foreground regions making use of two different strategies. Firstly, boundary information at the convolutional block level is preserved through residual skip connections comprised of a $1\times 1 $ convolution and a batch normalisation operation, which have been depicted as ``Skip$\_$Conv (1 $\times$ 1) $+$ BN'' in Figure~\ref{newmethodblockdiagram}. For the structural information preservation, we have also employed identity skip connections between the encoder and the corresponding decoder blocks. These are shown by black lines with arrowheads in Figure~\ref{newmethodblockdiagram}. Our motivation to use identity skip connections as an alternative to dense skip pathways also stems from the notion that feature preservation within each convolutional block can help bridging the semantic gap between the encoder and decoder while helping to maintain computational overhead under check.

\begin{figure}[!b]
  \centering
  \includegraphics[width=1\textwidth]{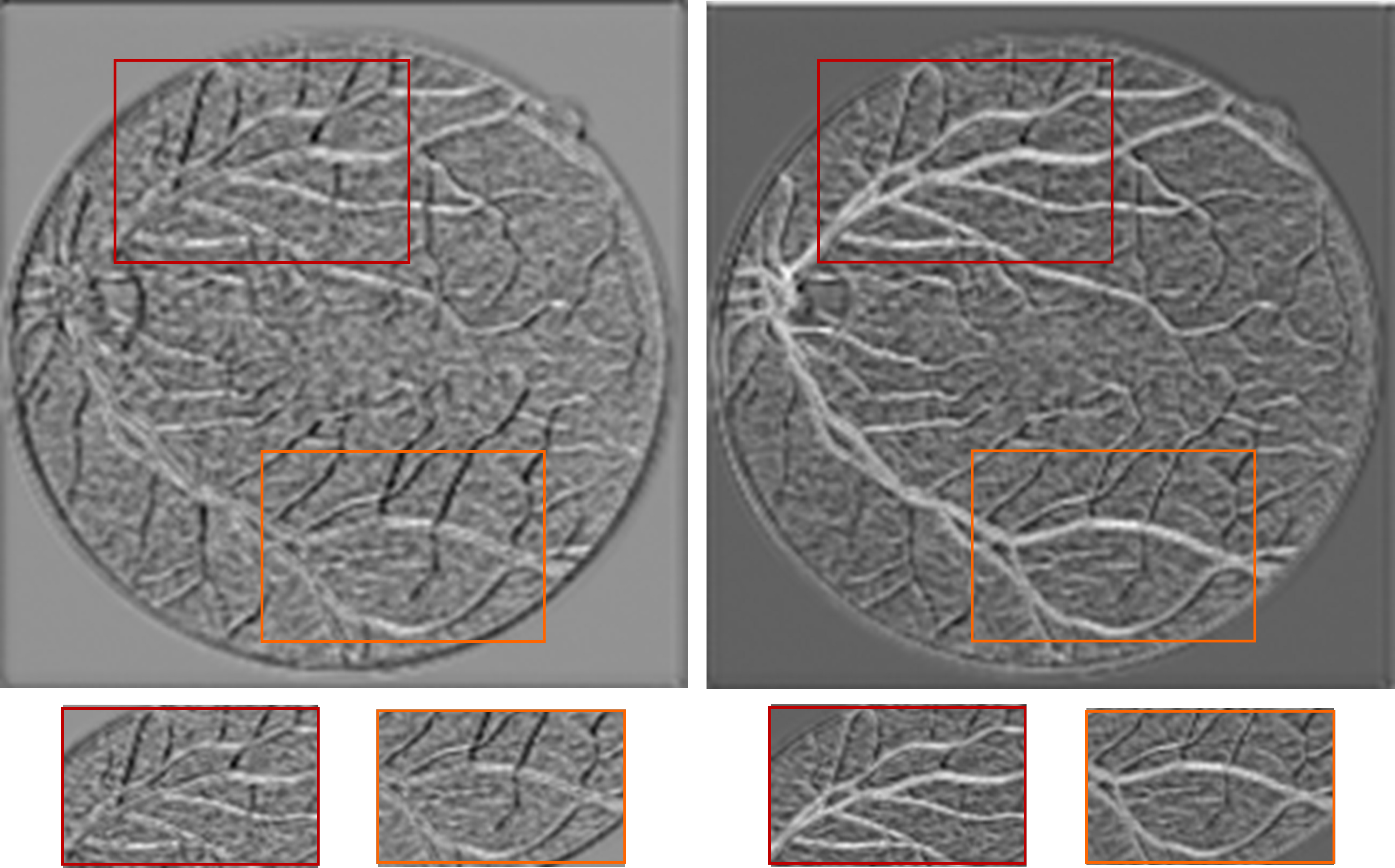} \\
  \caption{The structure preservation function of the residual skip connection within each convolutional block. The left and the right images show activation outputs before and after the addition of residual information.}
  \label{skip}%
\end{figure}%

Figure~\ref{skip} illustrates the effectiveness of our strategy as applied to boundary structure preservation. The left image shows a sample activation map obtained at the output of the third encoder block. Whereas, the right image shows the activation output after adding the residual information to the third encoder block output. For both images, we show the detail of the regions enclosed by red boxes under the activation map. It can be clearly observed from the details in the figure that much of the structural information is restored after the addition of the residual information.

\begin{figure}[!b]
  \centering
  \resizebox{1\textwidth}{!}{%
  \begin{tabular}{ccc}
       \includegraphics[width=0.5\textwidth]{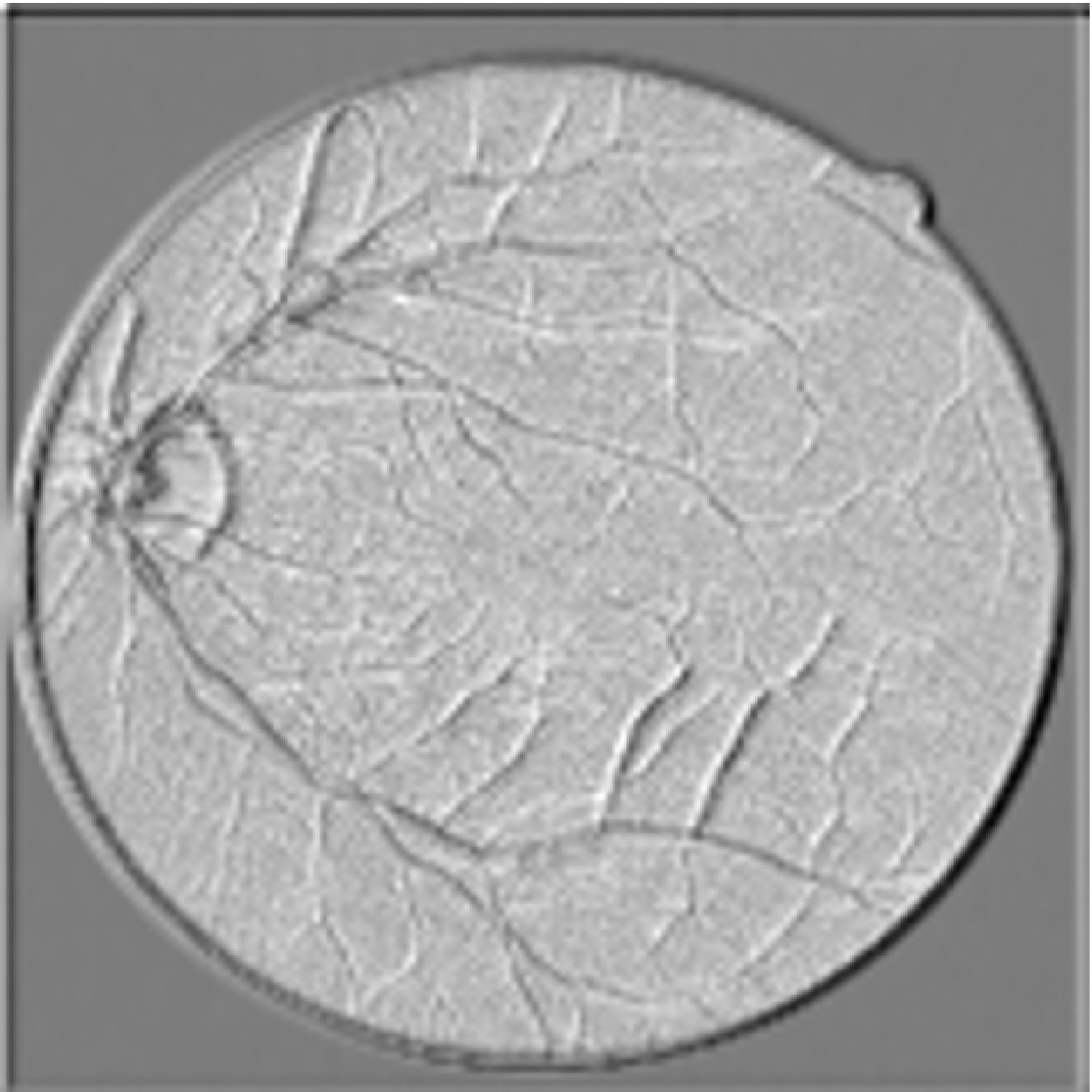}   &
       \includegraphics[width=0.5\textwidth]{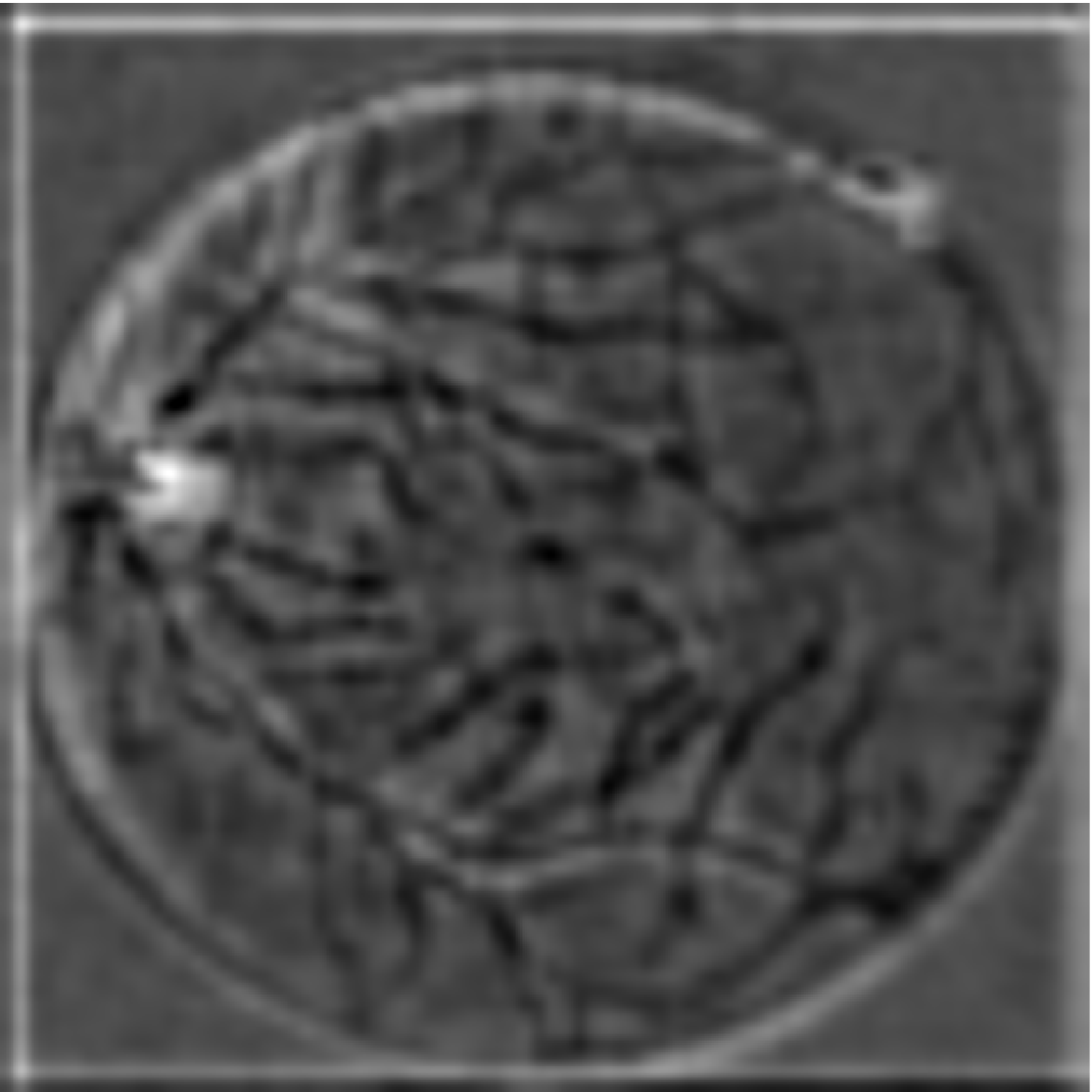} &
       \includegraphics[width=0.5\textwidth]{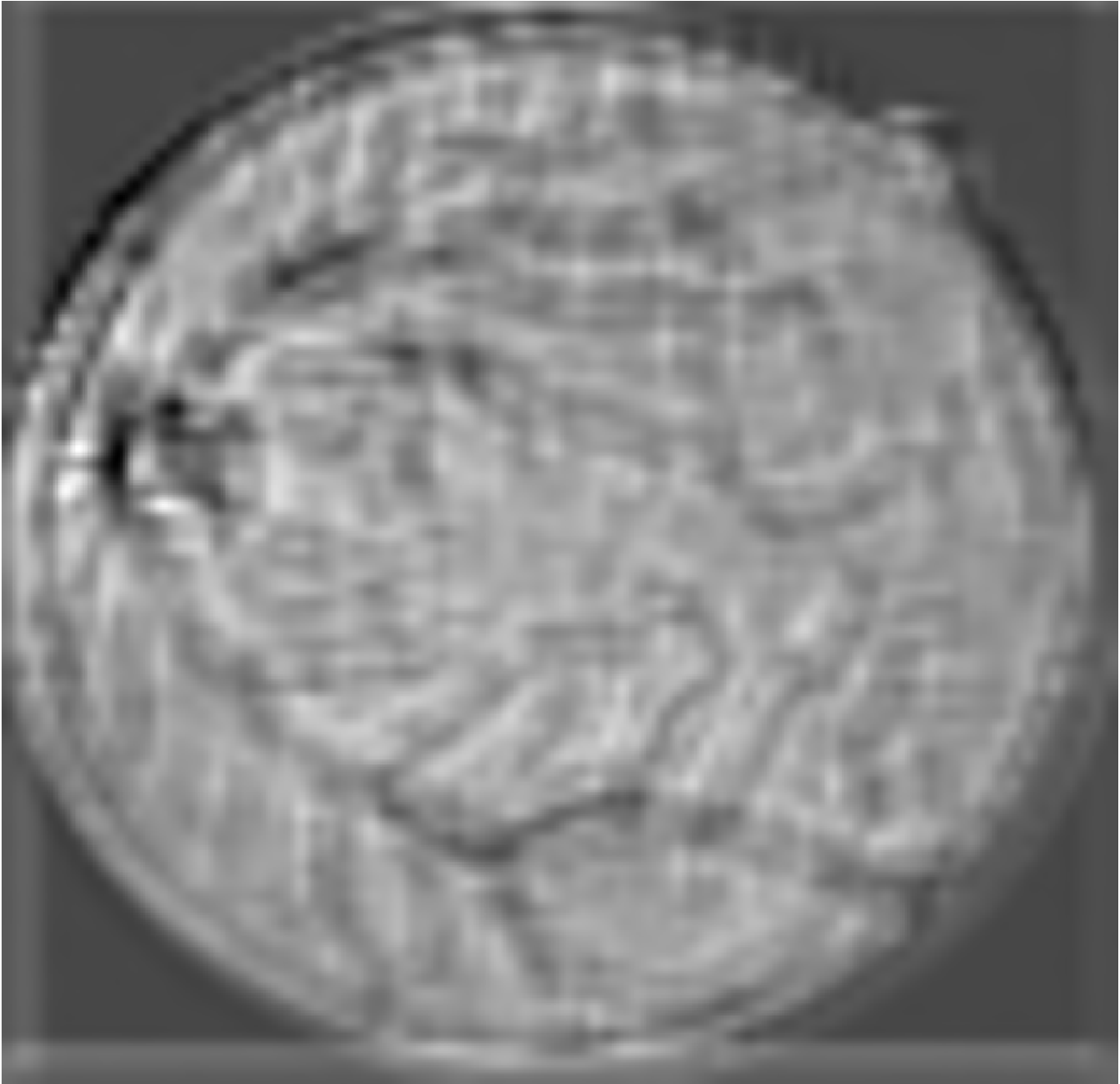} \\
  \end{tabular}
  }
\caption{Lowest resolution activation maps for RC-Net, SegNet and U-Net.}
  \label{pool}%
\end{figure}

We have also aimed at preserving this fine-grained structures, which often play a key part in medical image segmentation, by limiting the application of pooling operations in the network. Figure~\ref{pool} compares the lowest resolution activation of our network with the lowest resolution generally employed by the state-of-the-art encoder-decoder networks. In the figure, again, we have used a retinal image whereby the aim of computation is the segmentation of the vasculature. From left-to-right, the image shows the activation map for our network, SegNet\cite{M.Khan2020} and U-Net \cite{guo2020dpn}. Note that our network's map better preserves the spatial information as compared to those from the alternatives in the figure. All the activations were taken from the middle-layer of the networks under consideration, just before the first up-sampling block in each of the three architectures under consideration.

\section{Experimental Setup}\label{experimentalResults}
\subsection{Datasets}

For the retinal vessel segmentation, we have evaluated our network using two publicly available image data sets: DRIVE \cite{Staal2004}\footnote{The data set is widely available at \url{https://drive.grand-challenge.org/}} and STARE \cite{Hoover2000}\footnote{More information regarding the STARE project can be found at \url{https://cecas.clemson.edu/~ahoover/stare/}}. The DRIVE data set originated from a diabetic retinopathy screening program held in the Netherlands. It covers a wide age range of diabetic patients and consists of 20 color images for training and 20 color images for testing that are saved in JPEG format with an image size of 584$\times$565 pixels. Among these 40 images, only seven images have small signs of mild early diabetic retinopathy. A binary field of view (FOV) mask for each image is available. Both training and test images are equipped with manual vessel segmentation as ground truth that has been annotated by experts.

The STARE data set is a collection of 20 color retinal fundus images captured at $35^\circ$ FOV with an image size of 700$\times$605 pixels. Out of these 20 images, 10 images contain pathologies. Two different manual segmentation maps are available for each of these images. Here we employ the first expert segmentation as ground truth.

\subsection{Implementation and Training}


All our experiments have been effected on an Intel(R) Xeon(R) W-2133 3.6 GHz CPU with 96GB RAM and a GeForce GTX2080TI GPU. Our implementation of RC-Net used stochastic gradient descent with a fixed learning rate.

For all our experiments, a weighted cross-entropy loss is used as an objective function for training. This choice stems from the observation that, in vessel segmentation, the ``non-vessel'' pixels in each retinal image heavily outweigh the ``vessel'' pixels. For the assignment of the loss weights, different methods can be used. Here, we calculate class association weights by using median frequency balancing \cite{Badrinarayanan2017}.

Note that there is no dedicated test set available for STARE. For STARE, in the literature, typically a ``leave-one-out'' approach is used \cite{Soares2006}. Here, we have used both ``leave-one-out'' and a 50\%-50\% data split, i.e. 10 images for training and 10 for testing.


Also, since the retinal vessel segmentation data sets used here are quite small in nature, we have used data augmentation to generate enough data for training. For the data augmentation, we have used rotation and contrast enhancement. For the rotations, each training image is rotated by 1 degree. The contrast enhancement has been done by randomly increasing and decreasing the image brightness. This yields 7600 images for the DRIVE and 7000 images for each of the leave-one-out trails of the STARE data.




\subsection{Evaluation Criteria}

Recall that vessel segmentation maps are binary, whereby a pixel is marked as corresponding to a vessel or the background. The ``ground truth'' provided with publicly available datasets are manually marked by expert clinician. Thus, in each image, each pixel is classified into an area of interest (retinal vessels, skin lesions, intestinal polyps, etc.) if present. Note that, for each output image, there are four results: pixels which are correctly predicted as an area of interest (true positive (TP)), pixels which are correctly predicted as non-interesting (true negative (TN)),  non-interesting pixels incorrectly predicted as such (false positive (FP)) and, finally, interesting pixels incorrectly predicted as such (false negative (FN). Making use of this ingredients, in the literature, four common parameters (Sensitivity, Specificity, Accuracy and F1) are often used to compare methods with one another.

In the equations above the $Acc$ represents accuracy, showing the ratio of the pixels segmented correctly to the total number of pixels in the expertly annotated mask. The $Se$ and $Sp$ reflect the sensitivity and specificity that show how the vessel and non-vessel pixels are identified accurately in the model. In our vessel segmentation results we also show the area under the curve (AUC) for the receiver operating characteristic (ROC). We have done this since these data sets have an unbalanced distribution of positives and negatives \cite{Li2016}, where the AUC-ROC is often considered to be a good indicator of how the model can separate positive and negative classes in segmentation problems.

\begin{figure*}[htbp]
  \centering
  \begin{tabular}{ccccc}
       \includegraphics[width=0.14\textwidth]{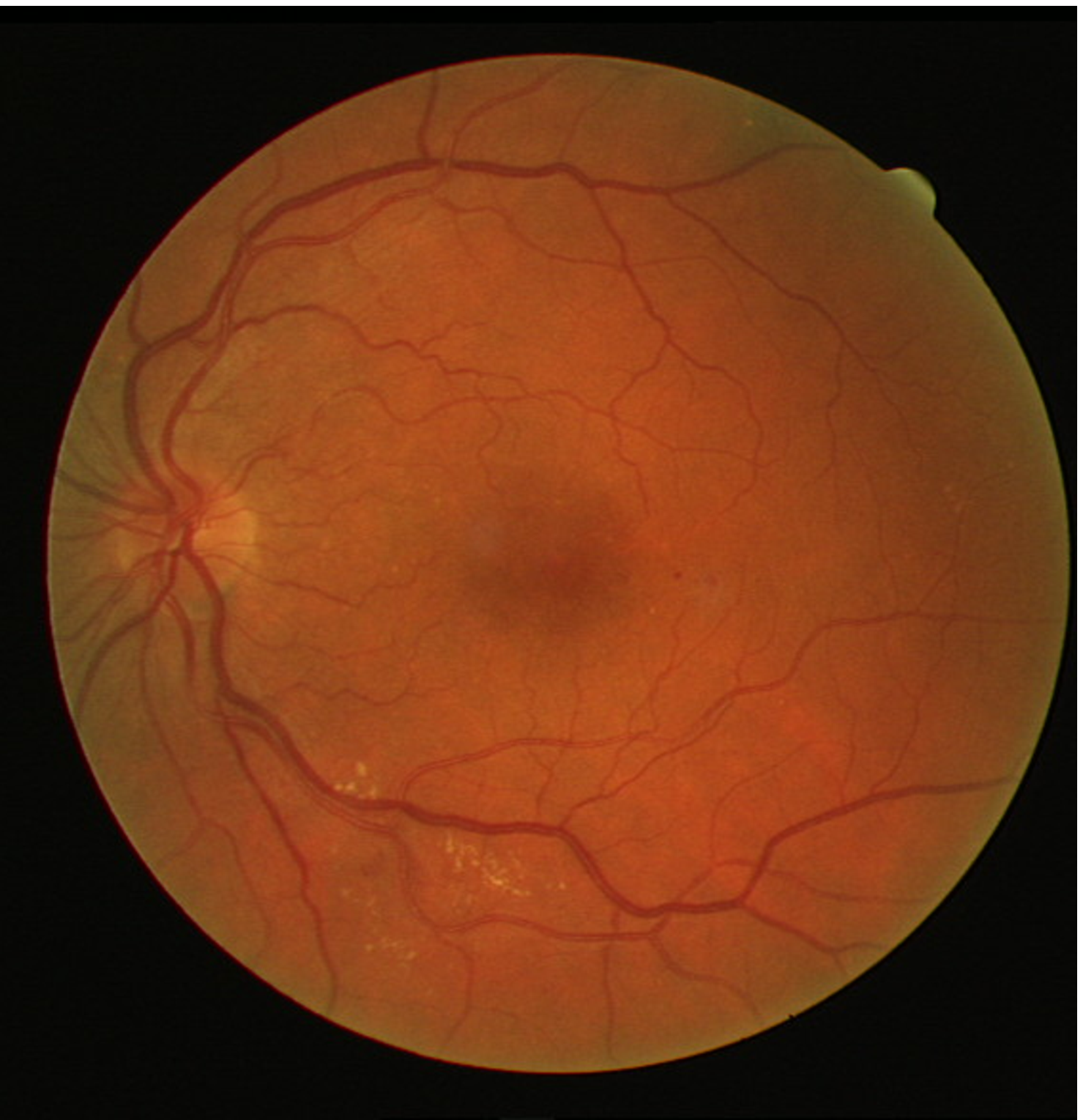}   & \includegraphics[width=0.14\textwidth]{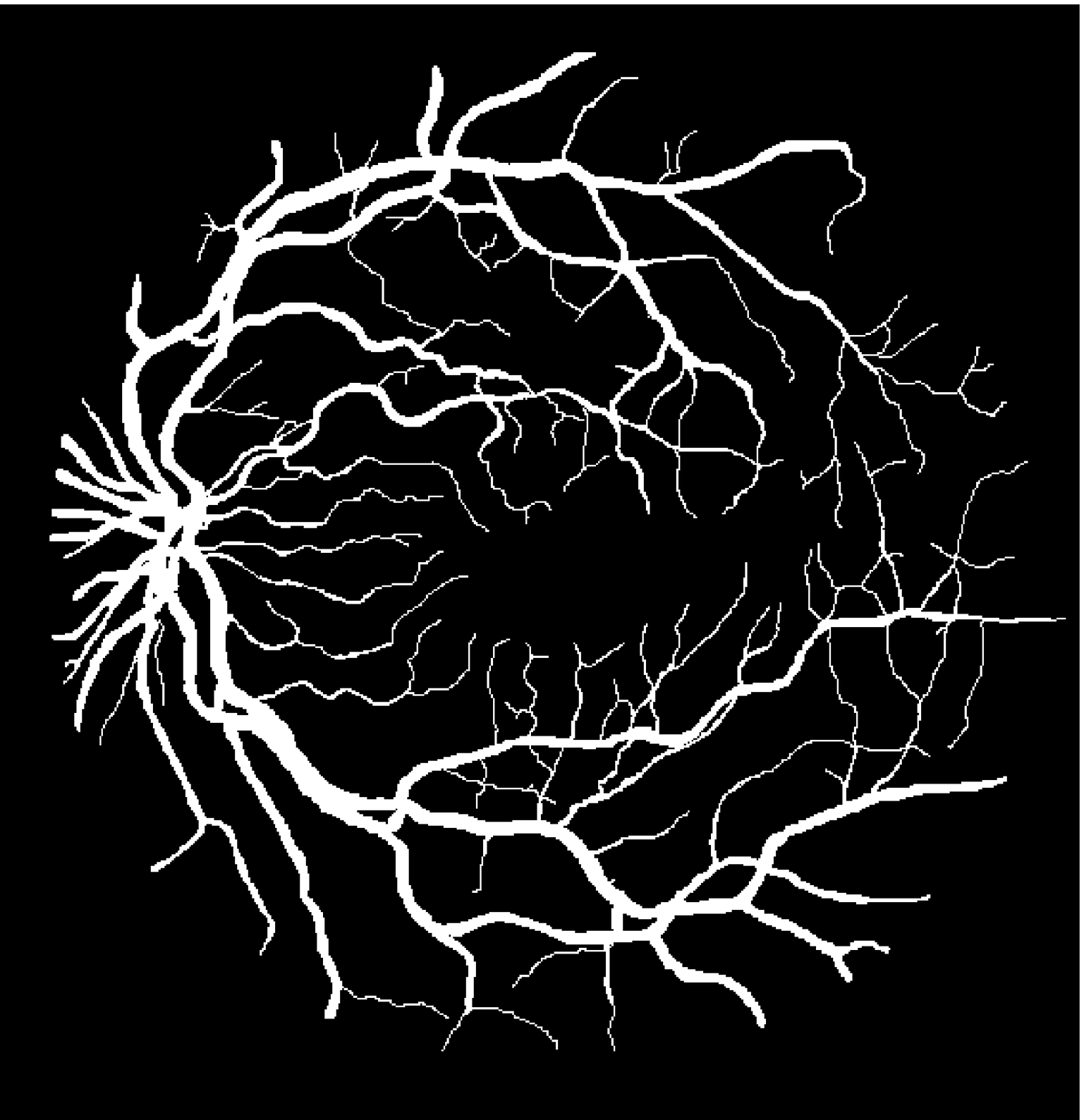}    & \includegraphics[width=0.14\textwidth]{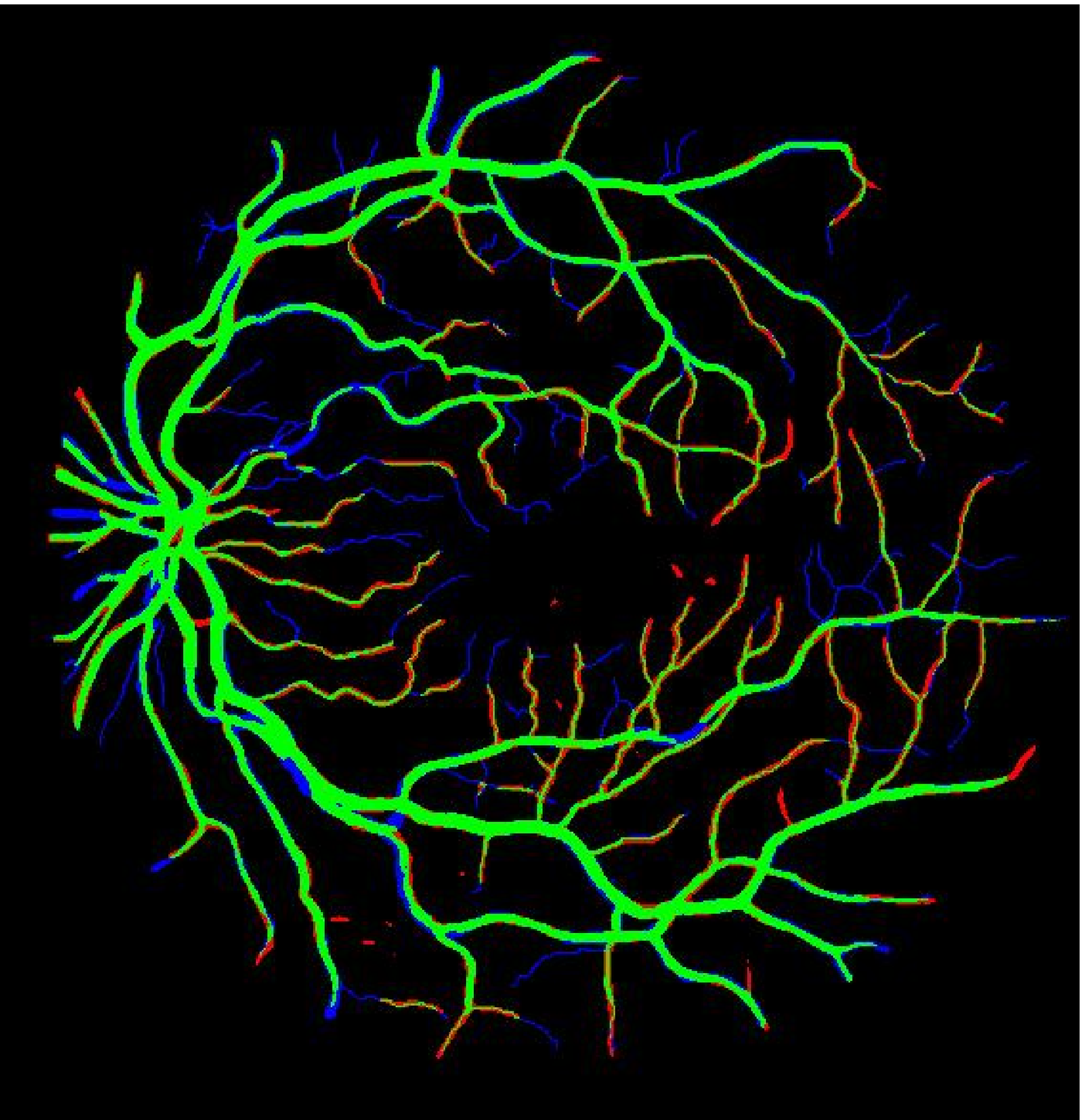}       & \includegraphics[width=0.14\textwidth]{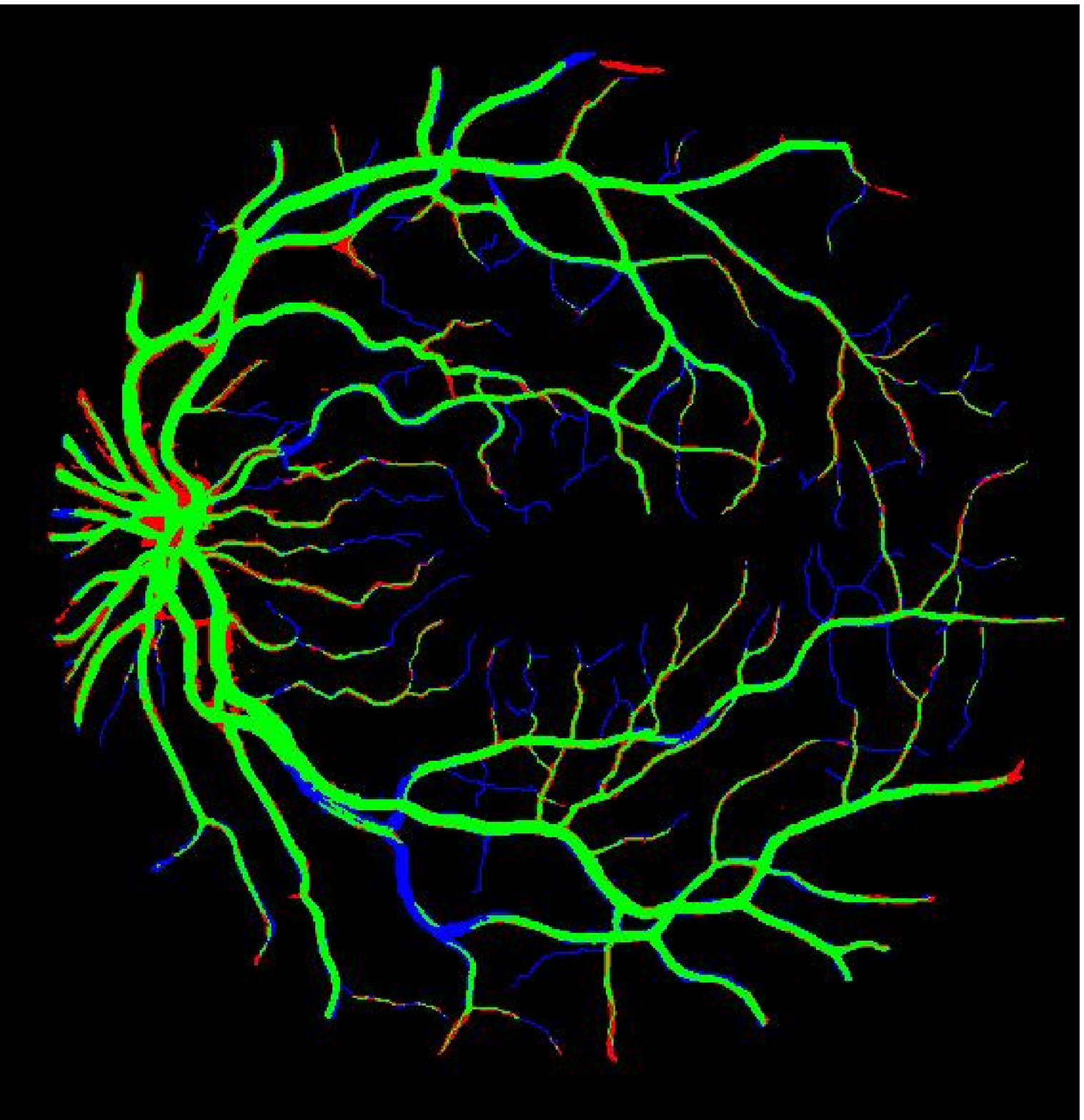}       & \includegraphics[width=0.14\textwidth]{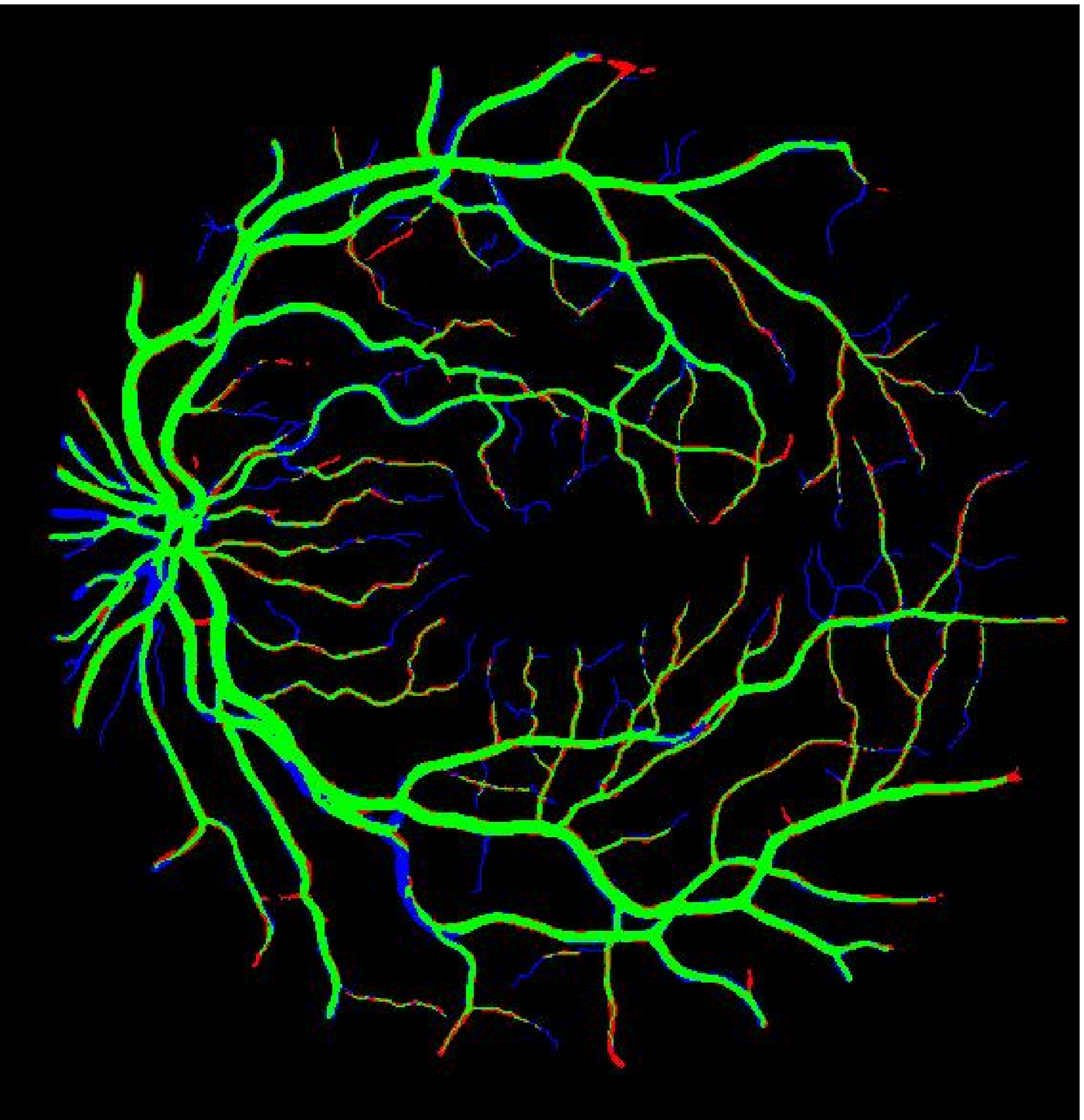}  \\
       \includegraphics[width=0.14\textwidth]{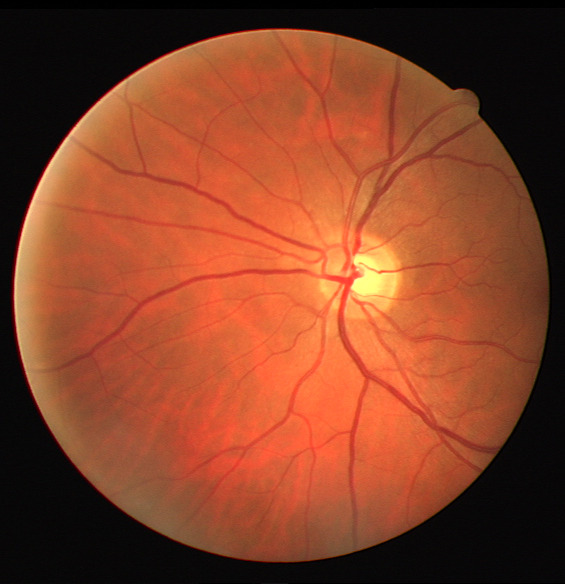}   & \includegraphics[width=0.14\textwidth]{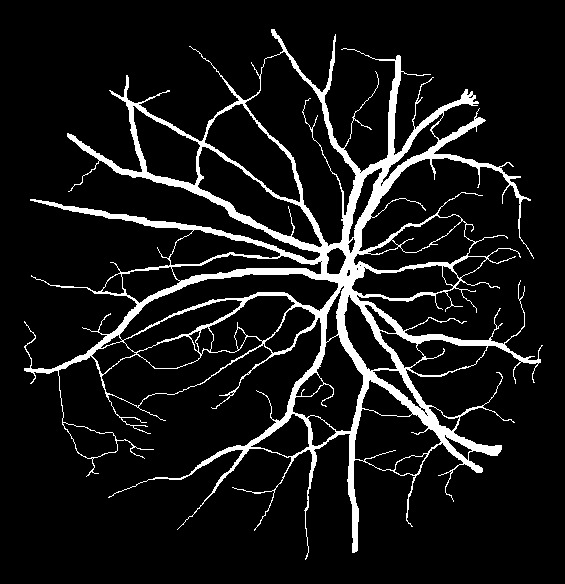}      &     \includegraphics[width=0.14\textwidth]{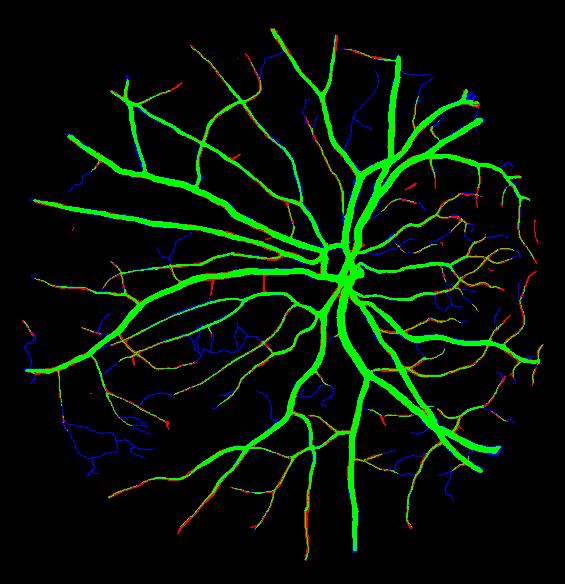}       & \includegraphics[width=0.14\textwidth]{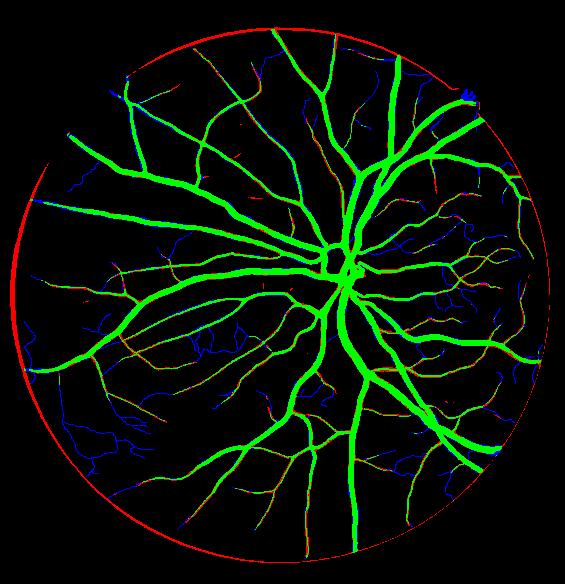}       & \includegraphics[width=0.14\textwidth]{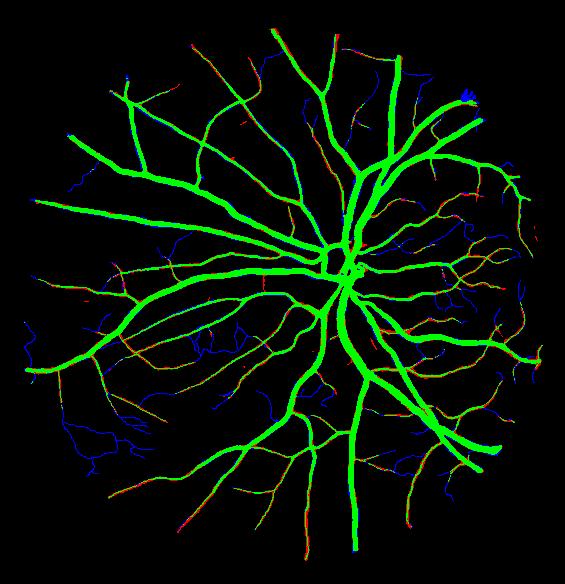}      \\
       \includegraphics[width=0.14\textwidth]{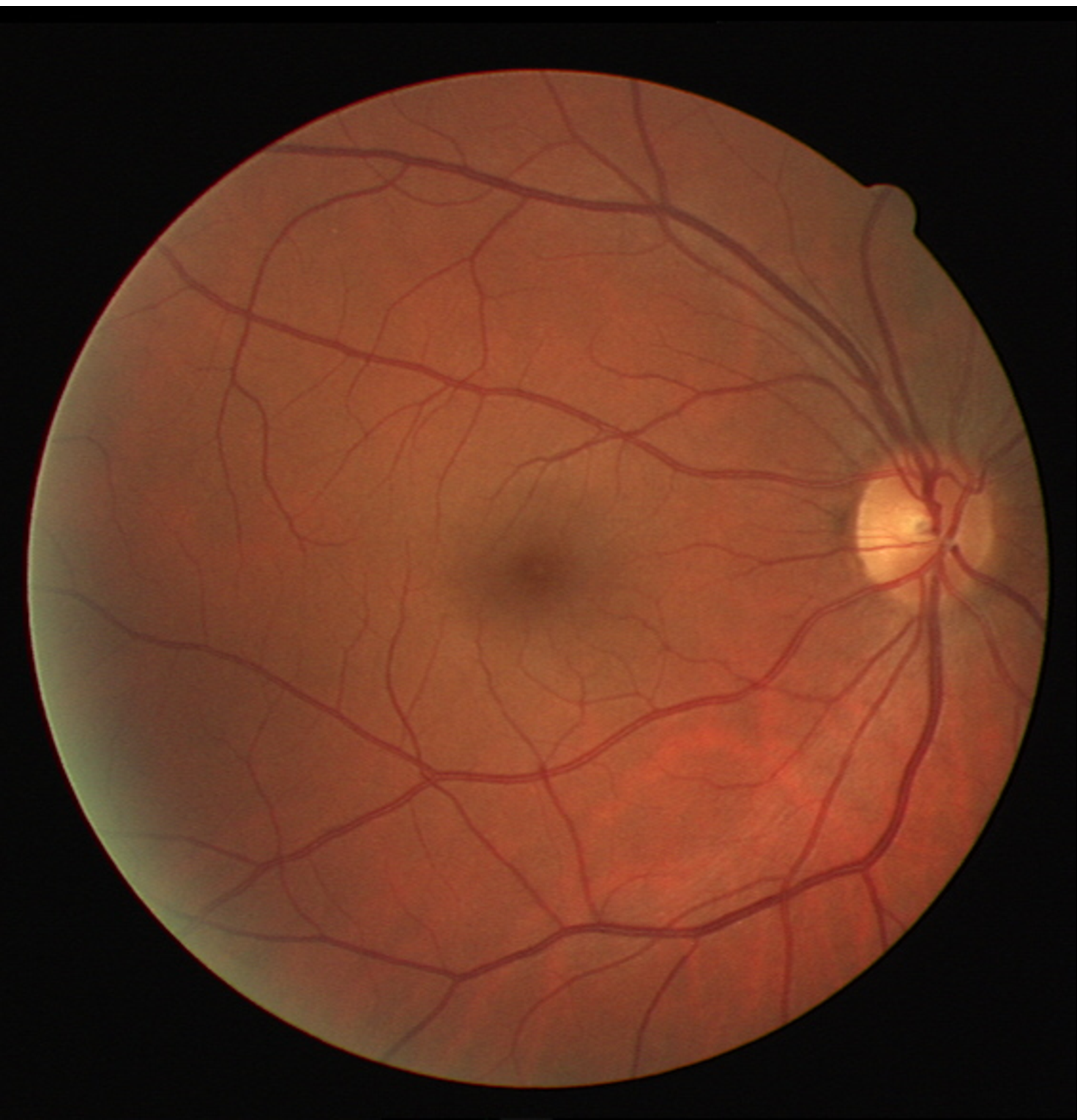}   & \includegraphics[width=0.14\textwidth]{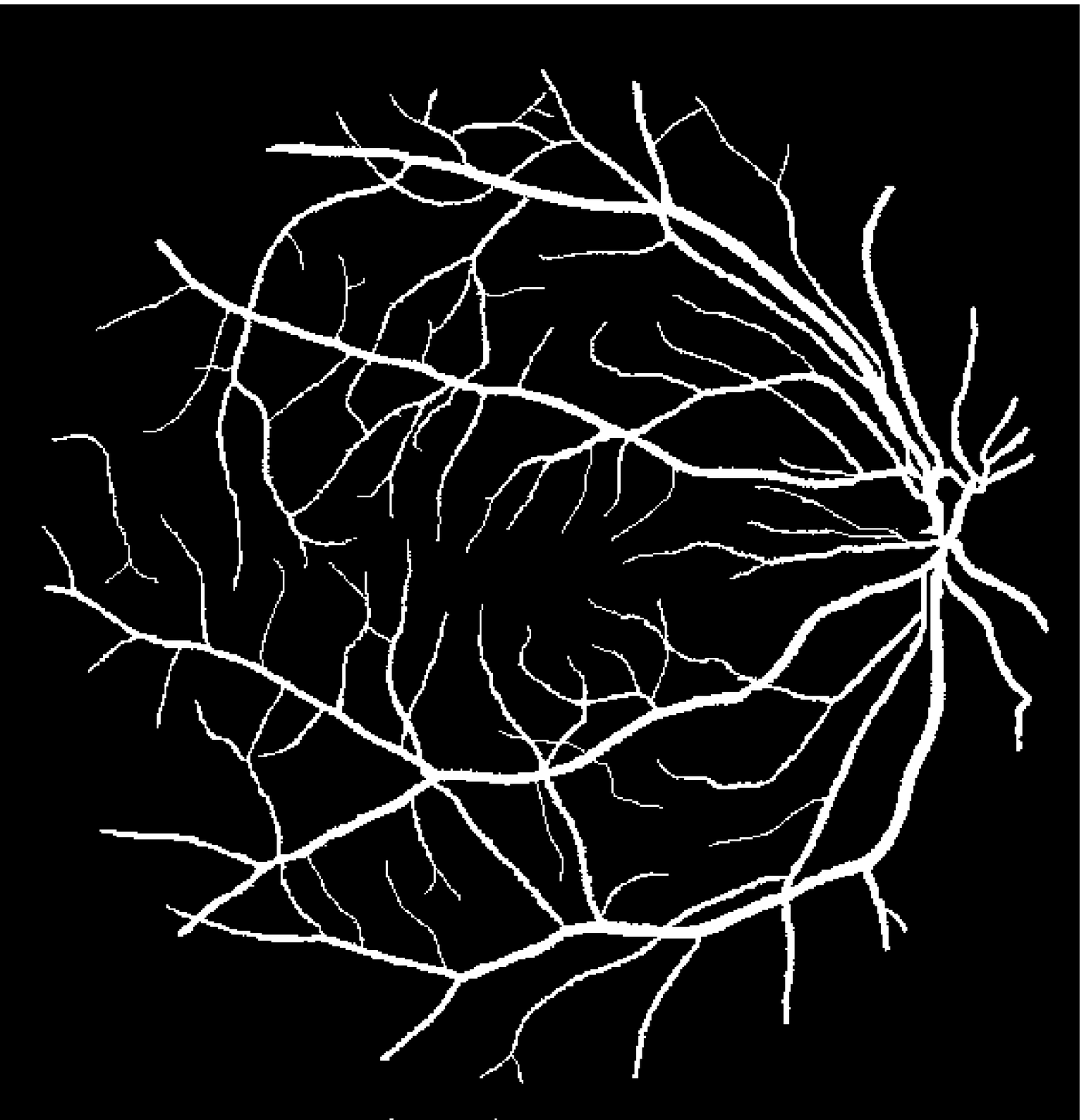}      &     \includegraphics[width=0.14\textwidth]{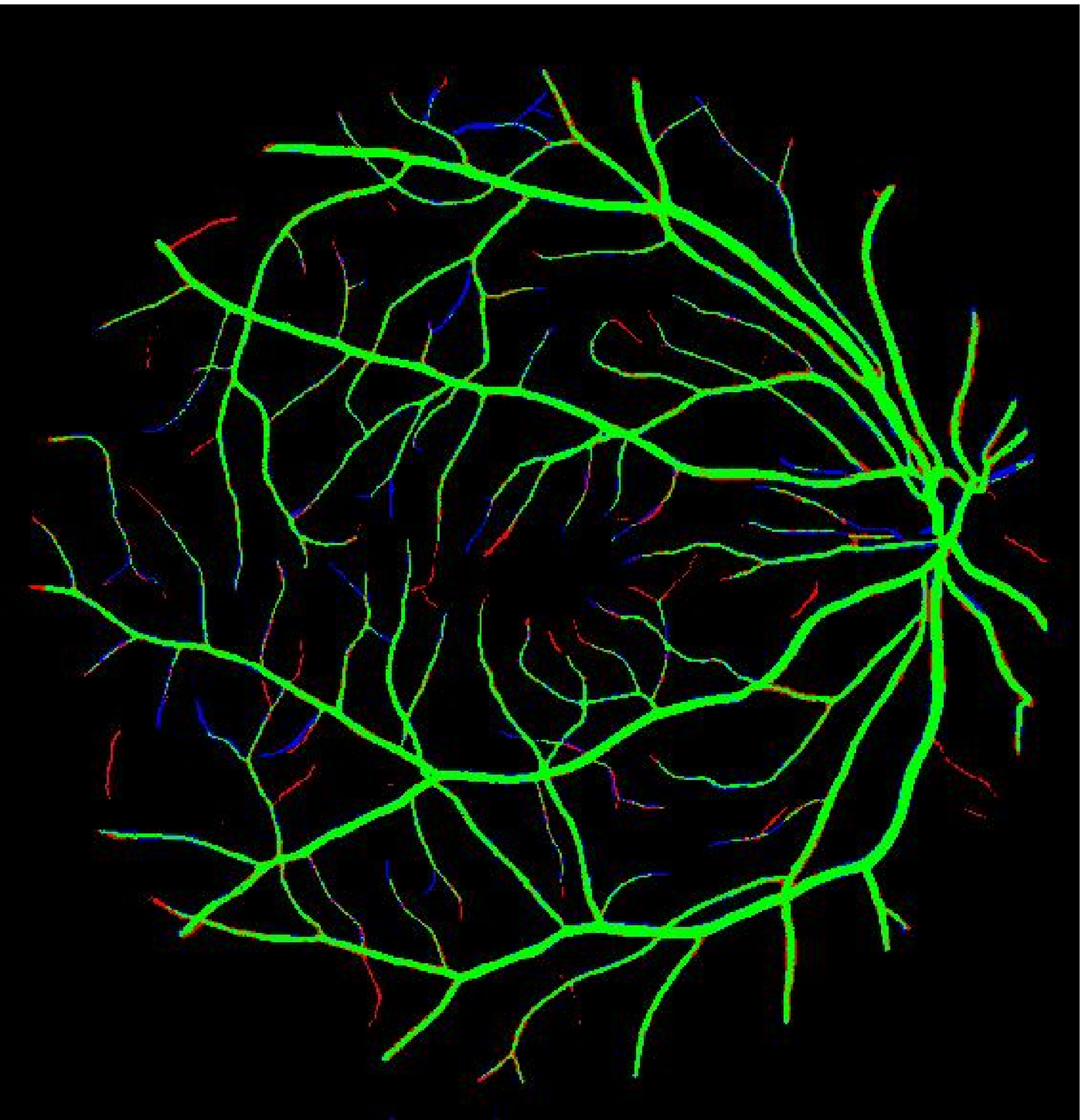}       & \includegraphics[width=0.14\textwidth]{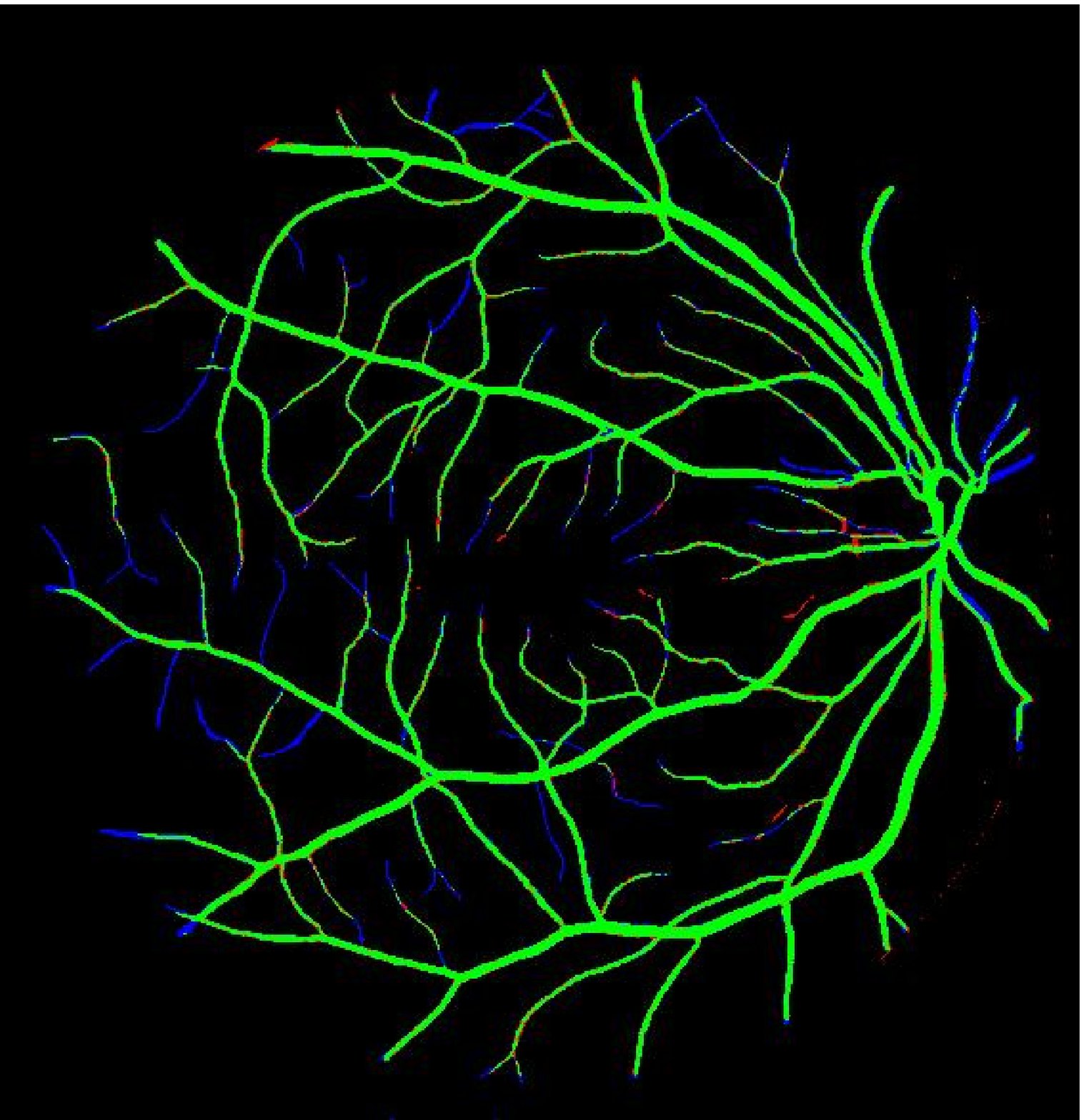}       & \includegraphics[width=0.14\textwidth]{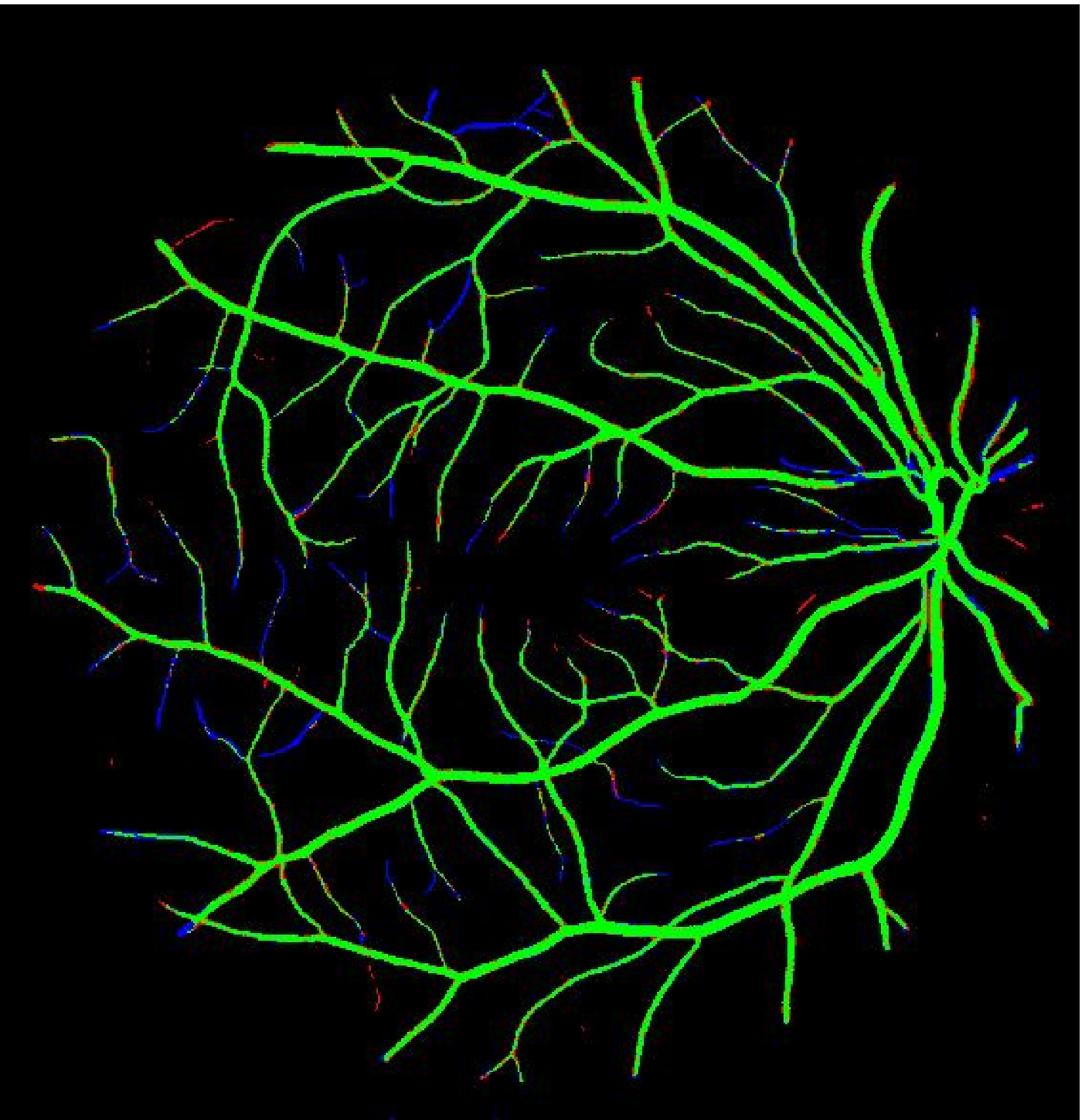}      \\
    \end{tabular}
    \caption{Sample retinal vessel segmentation results delivered by our network (RC-Net)  and two alternatives on the DRIVE dataset. From left-to-right, we show the input images, the ground truth vessel map manually annotated by an expert and the results yielded by SegNet \cite{M.Khan2020}, U-Net \cite{guo2020dpn} and RC-Net. The green and black colors in the segmentation maps show correctly predicted pixels whereas red and blue color pixels account for false positives and false negatives, respectively.}
  \label{visualDRIVE}%
\end{figure*}%

\begin{table*}[htbp]
  \centering
  \caption{Comparison results on the DRIVE dataset.}
    \begin{tabular}{ccccccc}
    \toprule
    \textbf{Method} & \textbf{Se} & \textbf{Sp} & \textbf{Acc} & \textbf{AUC} & \textbf{F1 score} & \textbf{Params(M)} \\
    \midrule
    SegNet \cite{M.Khan2020} & 0.7949 & 0.9738 & 0.9579 & 0.9720 & 0.8182   & 28.4 \\
    Three-stage FCN \cite{8476171} & 0.7631 & 0.982 & 0.9538 & 0.9750 & N.A   & 20.4 \\
    VessNet \cite{Arsalan2019}  &   0.8022  & 0.9810  & 0.9655  &0.9820   & N.A & 9\\
    DRIU \cite{Maninis2016} & 0.7855 & 0.9799 & 0.9552 & 0.9793 & 0.8220 & 7.8 \\
    Patch BTS-DSN \cite{GUO2019105} & 0.7891 & 0.9804 & 0.9561 & 0.9806 & 0.8249 & 7.8 \\
    Image BTS-DSN \cite{GUO2019105} & 0.78  & 0.9806 & 0.9551 & 0.9796 & 0.8208 & 7.8 \\
    U-Net \cite{guo2020dpn} & 0.7849 & 0.9802 & 0.9554 & 0.9761 & 0.8175 & 3.4 \\
    Vessel-Net \cite{Wu2019} & 0.8038 & 0.9802 & 0.9578 & 0.9821 & N.A   & 1.7 \\
    MS-NFN \cite{Wu2018} & 0.7844 & 0.9819 & 0.9567 & 0.9807 & N.A   & 0.4 \\
    FCN \cite{OLIVEIRA2018229} & 0.8039 & 0.9804 & 0.9576 & 0.9821 & N.A   & 0.2 \\
    \textbf{RC-Net} & \textbf{0.8319} & \textbf{0.9826} & \textbf{0.9694} & \textbf{0.9864} & \textbf{0.8262} & \textbf{0.027} \\
    \bottomrule
    \end{tabular}%
  \label{DRIVE}%
\end{table*}%

\section{Results and Comparison}
\label{sct:results}
We present a quantitative and qualitative analysis of our network, as well as a number of alternatives commonly used in retinal image segmentation. To that end, we'll start with the DRIVE dataset's qualitative retinal vessel segmentation results. In the figure, we show, from left-to-right, the input images, the expertly annotated vessel map (ground truth) and the segmentation results yielded by our method, SegNet \cite{M.Khan2020} and U-Net \cite{guo2020dpn}. Green and black represent correctly predicted vessel pixels in the figure, while red and blue represent false positives and false negatives, respectively. In Table \ref{DRIVE}, we show the corresponding qualitative results, now also including alternatives such as Image BTS-DSN \cite{GUO2019105} and Vessel-Net \cite{Wu2019}. It's worth noting that our network outperforms all of the alternatives by a small margin, despite having a much smaller number of parameters. This is supported by our qualitative findings, which show that the three networks produce segmentation maps that are very similar.

\begin{figure*}[htbp]
  \centering
  \resizebox{1\textwidth}{!}{%
  \begin{tabular}{ccccc}
       \includegraphics[width=0.14\textwidth]{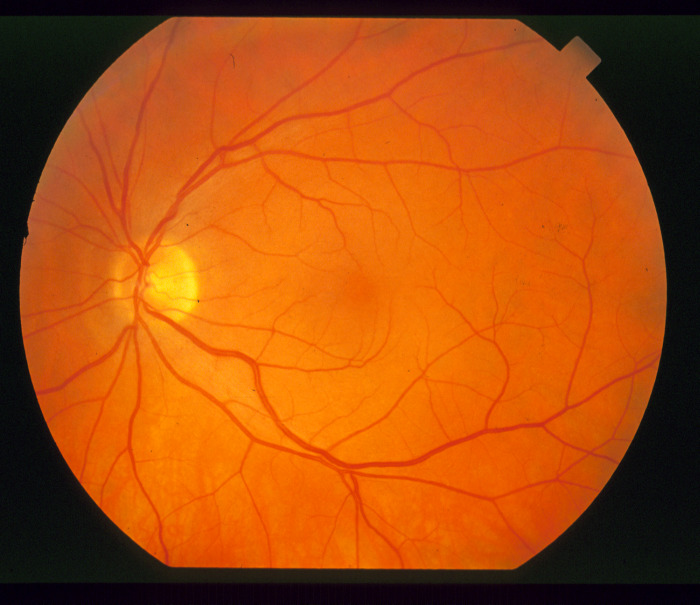}   & \includegraphics[width=0.14\textwidth]{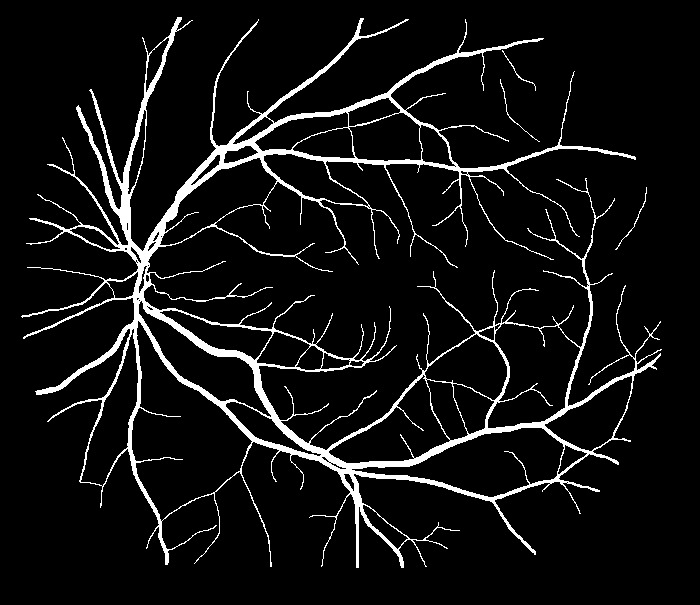}    & \includegraphics[width=0.14\textwidth]{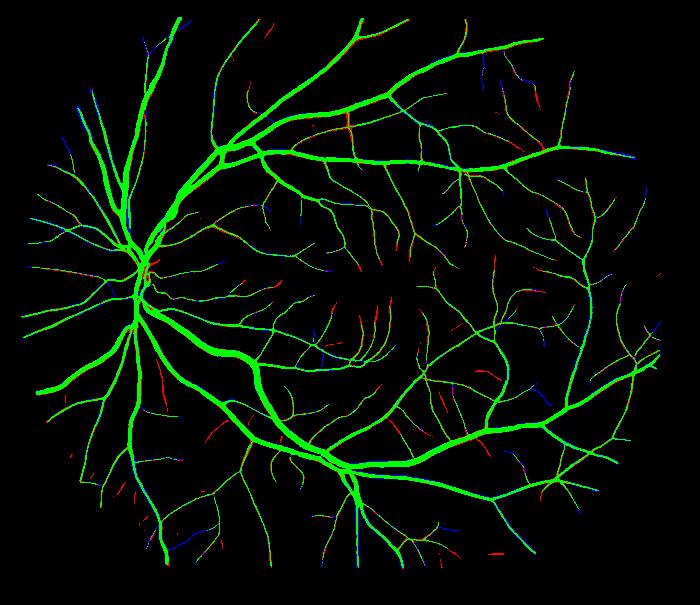}       & \includegraphics[width=0.14\textwidth]{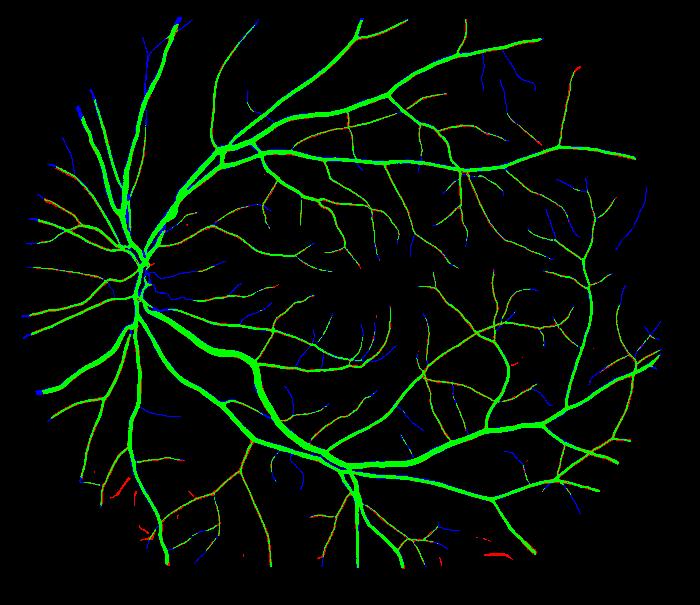}       & \includegraphics[width=0.14\textwidth]{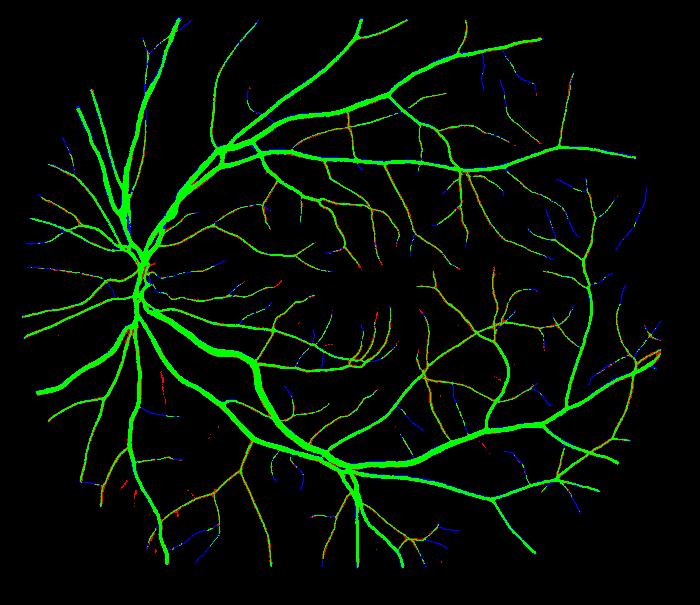}  \\
       \includegraphics[width=0.14\textwidth]{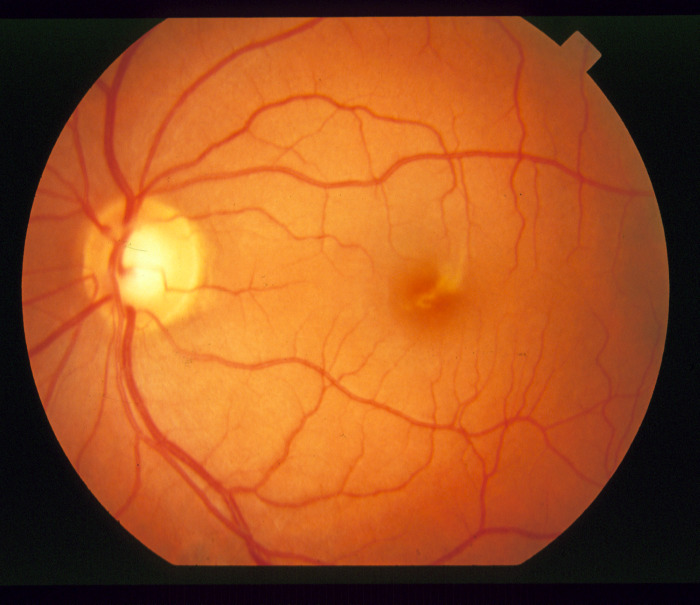}   & \includegraphics[width=0.14\textwidth]{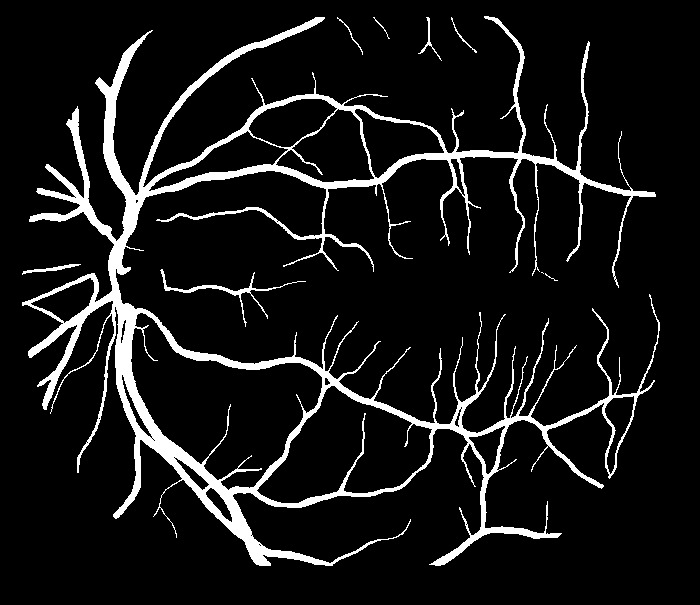}      &  \includegraphics[width=0.14\textwidth]{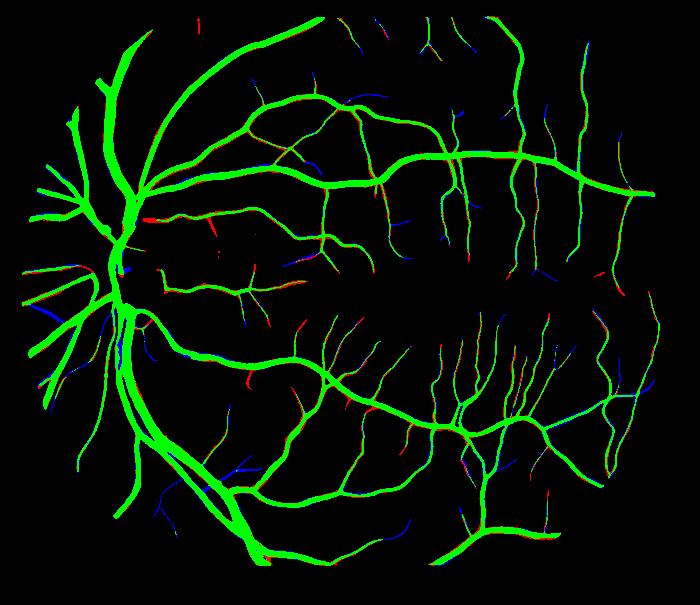}       & \includegraphics[width=0.14\textwidth]{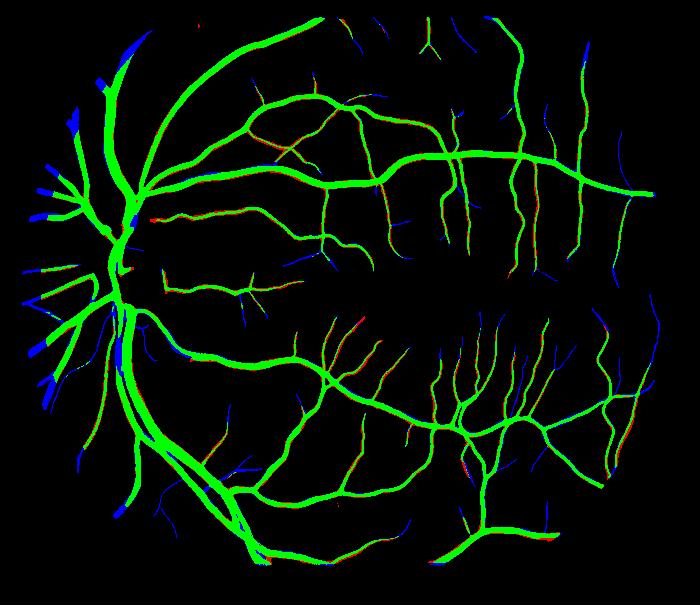}       & \includegraphics[width=0.14\textwidth]{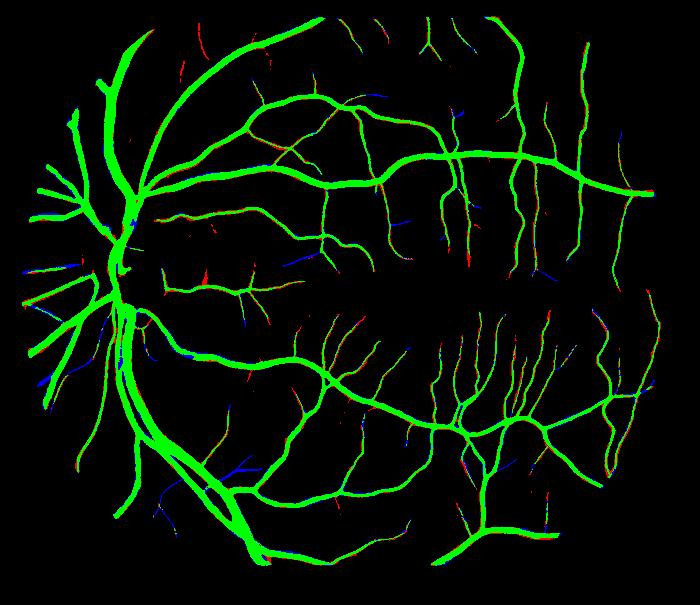}      \\
       \includegraphics[width=0.14\textwidth]{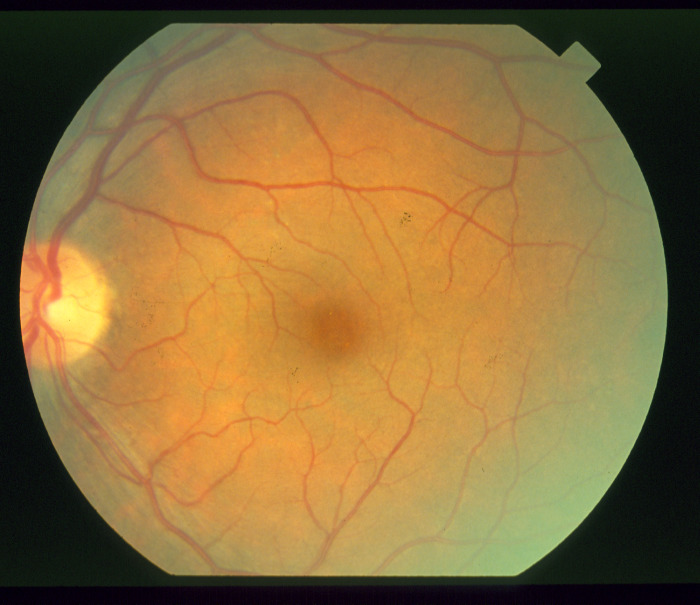}   & \includegraphics[width=0.14\textwidth]{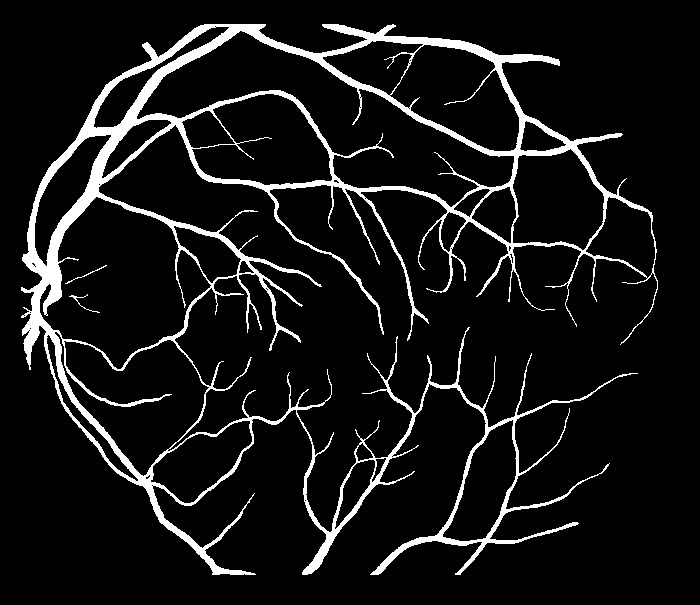}      &     \includegraphics[width=0.14\textwidth]{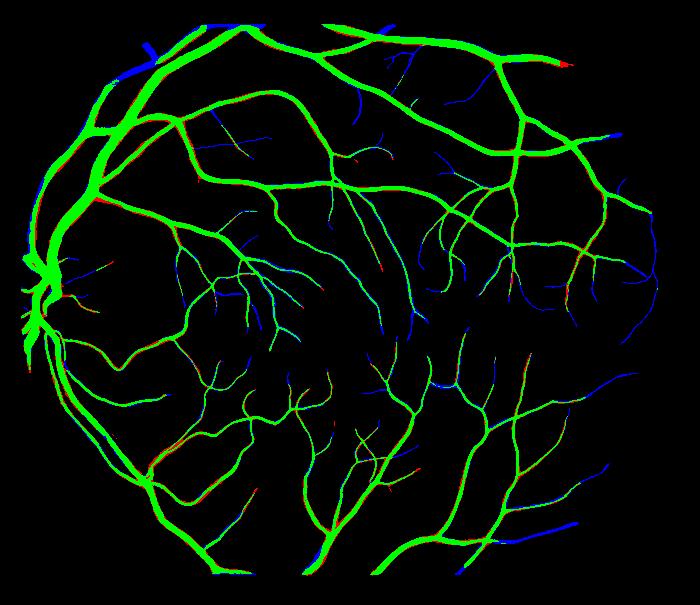}       & \includegraphics[width=0.14\textwidth]{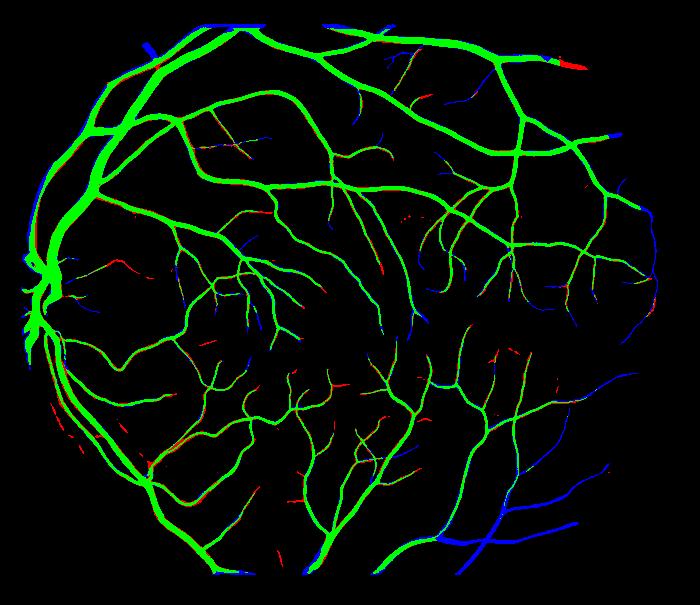}       & \includegraphics[width=0.14\textwidth]{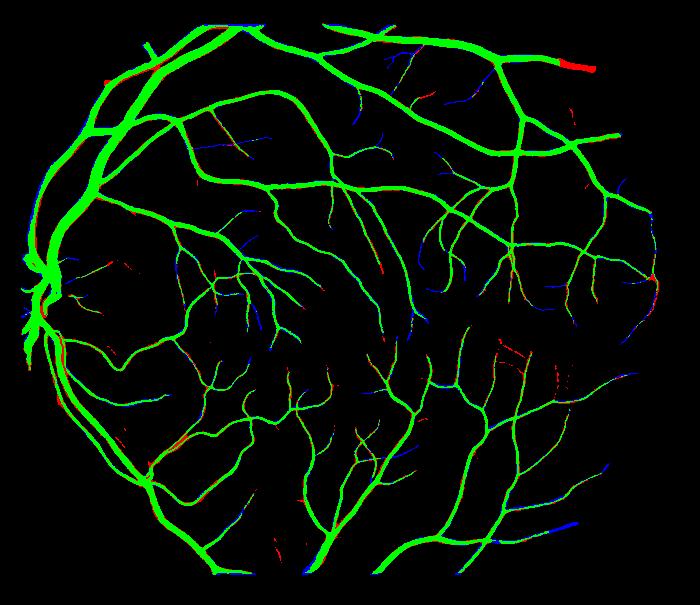}      \\
    \end{tabular}%
    }
    \caption{Sample retinal vessel segmentation results delivered by our network (RC-Net)  and two alternatives on the STARE dataset. From left-to-right, we show the input images, the ground truth vessel map manually annotated by an expert and the results yielded by SegNet \cite{M.Khan2020}, U-Net \cite{Ronneberger2015} and RC-Net. The green and black colors in the segmentation maps denote correctly segmented pixels. red and blue color pixels account for false positives and false negatives, respectively.}
  \label{visualSTARE}%
\end{figure*}%

\begin{table*}[htbp]
  \centering
  \caption{Results on the STARE dataset.  Best results are in bold.}
    \begin{tabular}{ccccccc}
    \toprule
    \textbf{Method} & \textbf{Se} & \textbf{Sp} & \textbf{Acc} & \textbf{AUC} & \textbf{F1 score} & \textbf{Dataset split} \\
    \midrule
    DRIU \cite{Maninis2016} & 0.8036 & 0.9845 & 0.9658 & \textbf{0.9970} & 0.831 & 50\%/50\% (train/test) \\
    Patch BTS-DSN \cite{GUO2019105} & 0.8212 & 0.9843 & 0.9674 & 0.9859 & \textbf{0.8421} & 50\%/50\% (train/test) \\
    Image BTS-DSN \cite{GUO2019105} & 0.8201 & 0.9828 & 0.966 & 0.9872 & 0.8362 & 50\%/50\% (train/test) \\
    SegNet \cite{M.Khan2020} & 0.8118 & 0.9738 & 0.9543 & 0.9728 & 0.8162 & 50\%/50\% (train/test) \\
    \textbf{RC-Net} & \textbf{0.8427}   & \textbf{0.9870}     & \textbf{0.9761}     & 0.9894     & 0.8410     & 50\%/50\% (train/test) \\
    \midrule
    U-Net \cite{guo2020dpn} & 0.764 & \textbf{0.9867} & 0.9637 & 0.9789 & 0.8133 & leave-one-out \\
    VessNet \cite{Arsalan2019}   &   \textbf{0.8526}   & 0.9791  & 0.9697 & \textbf{0.9883}& N.A& leave-one-out \\
    Three-stage FCN \cite{8476171} & 0.7735 & 0.9857 & 0.9638 & 0.9833 & N.A   & leave-one-out \\
    \textbf{RC-Net} & 0.8419 & 0.9858 & \textbf{0.9751} & 0.9881 & \textbf{0.8358} & leave-one-out \\
    \bottomrule
    \end{tabular}%
  \label{STARE}%
\end{table*}%

Now we'll take a look at the STARE data set. In Figure \ref{visualSTARE} we show sample results as delivered by our network, SegNet \cite{M.Khan2020} and U-Net \cite{guo2020dpn}. We show quantitative results in Table \ref{STARE}. Again, note that our method performs quite competitively against all the alternatives for both, leave-one-out and 50\%-50\% training strategies while being much more economical in terms of size as compared to all the other methods under consideration. For instance, the second best F1 score after RC-Net on leave-one-out is Patch BTS-DSN \cite{GUO2019105}, whose trainable parameters are almost 200 times as many as those in our network. U-Net \cite{guo2020dpn}, which has the best accuracy in leave-one-out has over 3.4 million parameters. RC-Net comes second with more than 100 times less parameters.

\section{Conclusions}

In this paper, we have presented RC-Net, a convolutional neural network for medical image segmentation which is quite small as compared to alternatives elsewhere in the literature.   
Here, we have kept pooling operations to a minimum and integrated skip connections into the network so as to preserve spatial information. Our network also employs a small number of kernels per convolutional layer. We have illustrated the utility of RC-Net in retinal vessel segmentation on four different datasets. In our experiments, RC-Net is quite competitive, outperforming a number of alternatives that are much larger in terms of trainable parameters.

{\small
\bibliographystyle{IEEEtran}
\bibliography{egbib}
}
\end{document}